# Direct detection of single molecules by optical absorption

M. Celebrano,[1+] P. Kukura,[1,2+] A. Renn,[1] V. Sandoghdar[1]*

[1] Laboratory of Physical Chemistry and *optETH*, ETH Zurich, 8093 Zurich, Switzerland.

[2] Current address: Laboratory of Physical and Theoretical Chemistry, Department of Chemistry, University of Oxford, Oxford OX1 3QZ, UK.

**Abstract**

The advent of single molecule optics has had a profound impact in fields ranging from biophysics to material science, photophysics, and quantum optics. However, all existing room-temperature single molecule methods have been based on fluorescence detection of highly efficient emitters. Here we demonstrate that standard, modulation-free measurements known from conventional absorption spectrometers can indeed detect single molecules. We report on quantitative measurements of the absorption cross section of single molecules under ambient condition even in their dark state, for example during photoblinking or strong quenching. Our work extends single-molecule microscopy and spectroscopy to a huge class of materials that absorb light but do not fluoresce efficiently

* To whom correspondence should be addressed: Email: vahid.sandoghdar@ethz.ch

[+] These authors contributed equally.

Ensembles of emitters are routinely studied via absorption and fluorescence alike, but absorption spectroscopy is extremely difficult to perform on very dilute samples because one has to detect very small changes on top of a large signal. Therefore, with the exception of cryogenic measurements (*1-3*), single molecule methods exclusively rely on the detection of fluorescence, where spectral filtering allows for background-free detection (*4*). The noise floor of this technique is, however, dictated by the detector dark counts so that it requires bright emitters to achieve a sufficiently high signal-to-noise ratio (*5*). Although single molecule fluorescence detection has brought about breakthroughs in optical imaging (*6*), bioanalytics (*7*), and fundamental photophysics (*8-10*), its critical dependence on highly efficient fluorescence has restricted its applicability to a tiny fraction of species in our surroundings. Several alternative contrast mechanisms such as photothermal (*11*) and interferometric interactions (*12-14*) have been explored to bypass this shortcoming for the optical detection of single nano-objects, but none has reached the single molecule level. In this report, we demonstrate that by reducing the interfering background and by proper optimization of the detection, it is possible to achieve shot-noise limited sensitivity below the parts-per-million level, and therefore, detect and study single molecules by absorption at room temperature.

Our experimental approach mimics a monochromatic version of an ultraviolet/visible spectrometer, where the absorption of a species is quantified by comparing the intensity of a light beam passing through a sample with and without the species of interest [see Fig. 1(a)]. To achieve this, we split the linearly-polarized output of a fiber-coupled helium-neon laser at a wavelength of 633 nm into probe and reference beams. The probe was injected into a home-built inverted microscope equipped with a closed-loop piezoelectric stage for sample scanning. The light was focused onto the sample and collected in transmission using two matched oil-immersion microscope objectives with numerical apertures 1.4. The probe and reference beams were focused onto a balanced photodetector, and their intensities were adjusted by neutral density filters for optimal common-mode noise rejection. Any fluorescence emitted by the sample was separated from the laser light by a dichroic mirror and additional long-pass filters and sent to an avalanche photodiode for single photon counting. A narrow-band filter was inserted in the probe path to reject any residual fluorescence. Our samples consisted of standard microscope cover glass coated by a thin polymer layer containing a nanomolar concentration of the molecular dyes terrylene diimide (TDI) (*15*) or Atto647N (Atto-tec, GmbH), both of which absorb efficiently at 633 nm [see Fig. 1(b)]. Index-matching at the polymer-oil interface was important for minimizing the interferometric background caused by the nanoscopic surface roughness (*16*).

Figure 2(a) depicts a fluorescence image obtained by raster scanning a TDI sample across the diffraction-limited focus spot of the laser. We observed significant variations in fluorescence intensity and spatial features caused by the interaction of linearly polarized light with randomly oriented dipole moments (*17*). To select the optimized projection of the illumination polarization onto the absorption dipole moment of the molecules under study, we performed our detailed measurements only on the most intense emitters for a constant incident power. After choosing a molecule, we recorded repeated lateral line scans across the maximum of its fluorescence spot [Fig. 2(b)]. The resulting fluorescence traces invariably exhibited fluorescence blinking and one-step photobleaching, which are well-established signatures of single molecules.

The simultaneously acquired absorption maps were dominated by shot-noise fluctuations of the laser intensity and residual interferometric scattering. Nevertheless, as shown in Fig. 2(c), in some cases a distinctive change in the differential transmission could be observed after photobleaching. Figure 2(d) displays fluorescence intensity and differential transmission averaged over 200 nm about the center of the molecule for each line in Figs. 2(b) and 2(c). The resulting traces reveal a stepwise change of about $3 \times 10^{-6}$ in the probe transmission occurring simultaneously with photobleaching.

The reproducibility of the features in Fig. 2(c) after photobleaching illustrates the residual contribution of sample surface roughness or slight variations in the index of refraction. The orange curve in Fig. 2(e) presents a quantitative measure of these signal fluctuations on the order of $3 \times 10^{-6}$ root-mean-square (RMS). Here, we have averaged successive line scans and thereby reduced the intensity fluctuations induced by the shot noise. The black trace in Fig. 2(e) shows that these background fluctuations could be suppressed further by an order of magnitude to $5.3 \times 10^{-7}$ RMS if two such averages were subtracted. The remaining fluctuations agree well with the expected shot-noise limit of $5 \times 10^{-7}$ for 100 μW incident power, 1 ms pixel dwell time, and averaging of 80 lines when optical losses (60%) and the quantum efficiency of the detector are accounted for (~50%).

Having quantified and minimized the fluctuations of the signal transmitted through the sample, we can now subtract two line scan averages chosen before and after photobleaching [see the arrows in Figs. 2(b) and 2(c)]. Figures 2(f) and 2(g) depict the results for the fluorescence and transmission signals, respectively. The averaged absorption scans during molecule emission (red trace) exhibit a clear transmission dip with a magnitude of $\Delta T/T \sim 4\times10^{-6}$ at the same lateral position as the fluorescence peak. We emphasize that as shown by the blue and grey traces in Fig. 2(g), no such dip could be observed if both averages were chosen from the regions before or after bleaching. The observation of a dip in the red curve of Fig. 2(g), therefore, confirms that the absorption cross section is diminished upon photobleaching. We note that the above-mentioned averaging and subtraction procedure allowed us to detect the absorption signal of single molecules even when they were not readily visible in the raw data, as was the case in Figs. 2(c) and 2(d).

A simple calculation using the molar ensemble extinction coefficient of 72000 $M^{-1}$ $cm^{-1}$ at 633 nm for TDI [see Fig. 1(b)] (*18*), yields an average cross section of about $3 \times 10^{-16}$ $cm^2$ for randomly oriented molecules. To obtain the cross section ($\sigma$) of a single molecule oriented parallel to the polarization of the incident light, this value has to be multiplied by a factor of three to give $9 \times 10^{-16}$ $cm^2$. If we now consider a diffraction-limited excitation spot with area A and equate the ratio $\Delta T/T$ to $\sigma/A$, our measurement on a TDI molecule lets us deduce $\sigma \sim 1.5 \times 10^{-15}$ $cm^2$. This value is very close to the quantity extracted from ensemble measurements, but we note that one should not expect a perfect agreement because the cross section can vary in different hosts. Furthermore, there could exist considerable variations from molecule to molecule, caused by the inhomogeneity in the absorption spectrum.

In Fig. 3(a) we report the results of measurements on more than thirty molecules. The linear correlation between the fluorescence and absorption signals confirms that the changes in $\sigma$ can be attributed to a distribution in the excitation efficiency and are not caused by measurement errors. Such variations might be caused by a residual orientation mismatch between the light polarization and the molecular absorption dipole moment or the inhomogeneous broadening of the absorption spectrum. Future simultaneous multicolour measurements (*12*) will give rise to single-molecule absorption spectra, thereby revealing this inhomogeneity directly. Although this phenomenon is known from cryogenic experiments (*19*), at room temperature only variations of emission spectra have been reported previously (*20*). We also remark in passing, that we were able to detect every molecule in absorption given that enough photons could be collected before photobleaching, and no sample drift occurred during the experiment.

In some cases our selection for maximum brightness resulted in two neighbouring molecules in the focus. The fluorescence line image in Fig. 3(b) and a time trace of the spatially averaged fluorescence intensity shown in Fig. 3(c) display two clear bleaching steps. The simultaneously acquired averaged absorption scans also revealed two dips in the differential transmission of different magnitudes for two molecules ($6.5x10^{-6}$) and one molecule ($3.6x10^{-7}$). This constitutes a further proof of the single molecule sensitivity achieved by our setup.

Our ability to probe the molecule even in the absence of fluorescence allowed us to investigate the absorption cross section of individual molecules during fluorescence blinking. The red and blue curves in Fig. 3(e) show the fluorescence line scans during the blinking on and off states, respectively. The red trace in Fig. 3(f) plots the line scan obtained by subtracting averages of the transmission signal when the molecule was emitting from those when it was intermittently switched off. The clear transmission dip at the same lateral position as the fluorescence maximum is evident. The blue trace in Fig. 3(f) confirms that subtraction of averages during the off-state do not show any feature at the position of the molecule. These results illustrate that the absorption cross section of a TDI molecule vanishes during the blinking off times because it is in a different electronic state. We note that this observation is in contrast to the behavior of single semiconductor nanocrystals where the extinction cross section remains unchanged during photoblinking (*16*).

A key advantage of absorption over fluorescence detection is its insensitivity to quenching. Here, it is important to remember that the fluorescence signal scales linearly with the quantum efficiency $\eta = \frac{\gamma_r}{\gamma_r + \gamma_{nr}}$, while the absorption dip is proportional to $\eta = \frac{\gamma_r}{\gamma_r + \gamma_{nr} + \gamma_{ph}}$. The parameters $\gamma_r$, $\gamma_{nr}$, and $\gamma_{ph}$ denote the radiative linewidth of the transition, the nonradiative decay of the excited state population, and the spectral broadening caused by the phononic interaction with the matrix. At room temperature, $\gamma_{ph} \sim 10^6\ \gamma_r$ so that even a quenching rate $\gamma_{nr}$ that is several orders of magnitude larger than $\gamma_r$ would not affect the absorption signal, while it fully diminishes the fluorescence. To demonstrate this phenomenon experimentally, we examined Atto647N molecules, which have been proposed to be quenched in polyvinyl alcohol (PVA) via an electron transfer process (*19*). Figure 4(a) confirms that the fluorescence of such a sample is extremely weak and undergoes intermittent emission. The blue curves of Figs. 4(b) and 4(c) display the averages of lines after photobleaching for fluorescence and absorption respectively, while the red traces in these figures plot the subtraction of averages before and after photobleaching. We again find an absorption dip at the same position as the fluorescence maximum. We note that although the emission level here is more than 30 times weaker than that of TDI, the absorption signal of a single Atto647N molecule embedded in PVA is still comparable to that of TDI. The lower value of the recorded absorption dip can be explained by our inability to confidently select well aligned molecules as well as the inevitable inclusion of blinking off-times in our averaged traces. The signal-to-noise ratio is inferior because of lower count rates, and thus, higher shot noise.

Since the physical mechanism underlying the extinction of a light beam by an emitter is interference (*2, 3*), our results demonstrate that direct interferometric sensing can reach single molecule sensitivity. Combination of this technique with specific surface functionalization (*21*) holds great promise to push optical biosensing to its limit without invoking microcavities (*22*) or plasmonics (*23*). In our current work, we have relied on the intrinsic photophysics of dye molecules (i.e. bleaching or blinking), to produce a very slow on-off signal modulation for separating the signal of interest from the background.

Similarly, it should be possible to employ high-speed modulation and lock-in detection (*24*) with the potential for further improvement in noise suppression. Importantly, measurements at different wavelengths open the door to absorption spectroscopy of single molecules which aside from its fundamental interest for studying molecule-host interactions would act as a chemical signature of the species under investigation.

ACKNOWLEDGEMENTS

This work was supported by ETH Zurich and the Swiss National Foundation. We thank Guido Grassi for the synthesis of TDI.

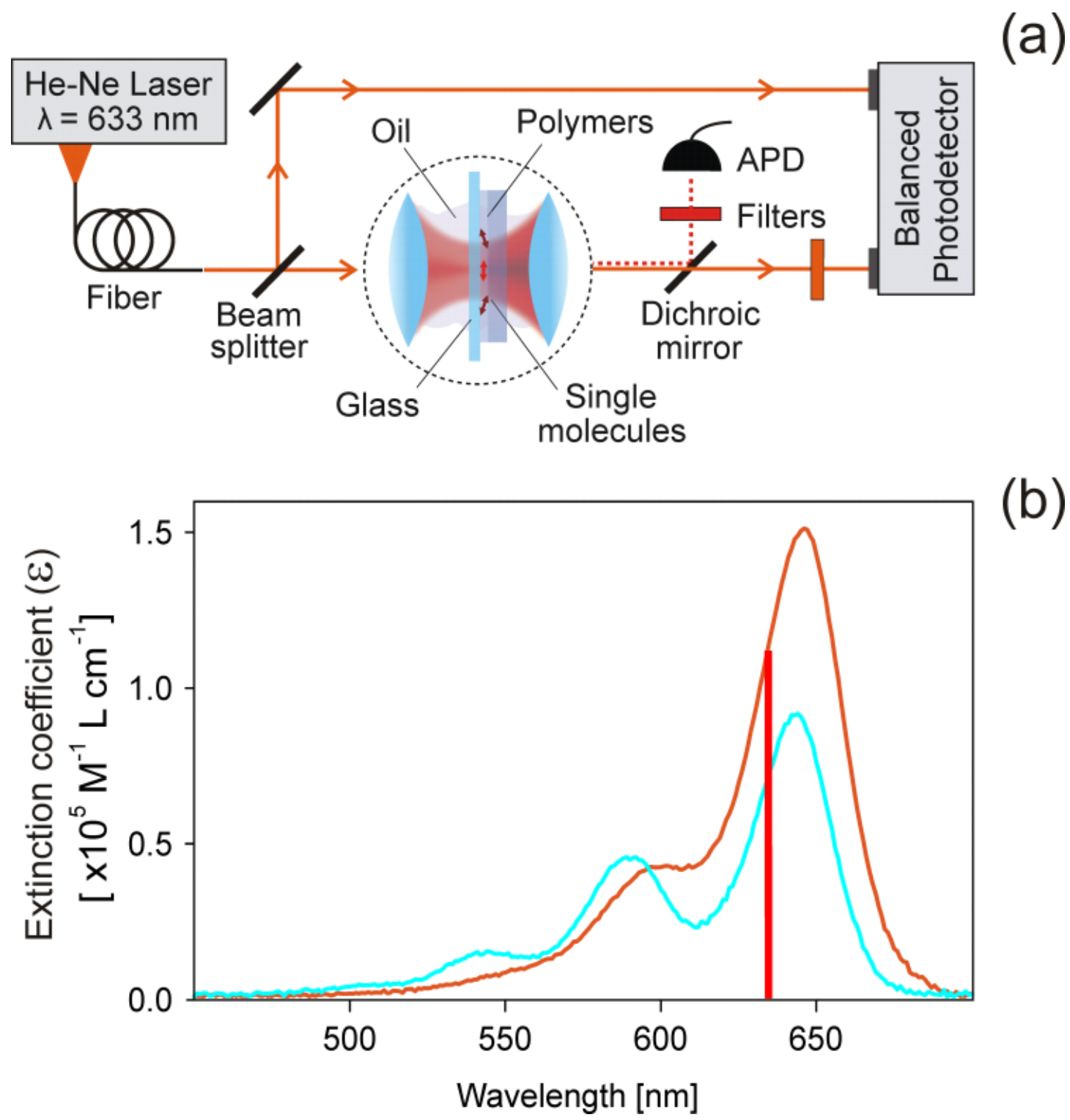

**Figure 1** (a) Experimental setup, APD: Avalanche photodiode, He-Ne: Helium-Neon. The single molecules studied were either terrylene diimide (TDI) or Atto647N. TDI was embedded in an 80 nm thin layer of polymethylmethacrylate (PMMA) and covered by a one micrometer layer of polyvinyl alcohol (PVA) for protection from microscope immersion medium. Atto647N was spin cast directly onto microscope cover glass and then protected by a 1 micron PVA layer. (b) Ensemble absorption spectra of TDI in toluene and Atto647N in water. The laser excitation wavelength at 633 nm is indicated for clarity. Note that these spectra may differ for the polymer hosts used in this experiment.

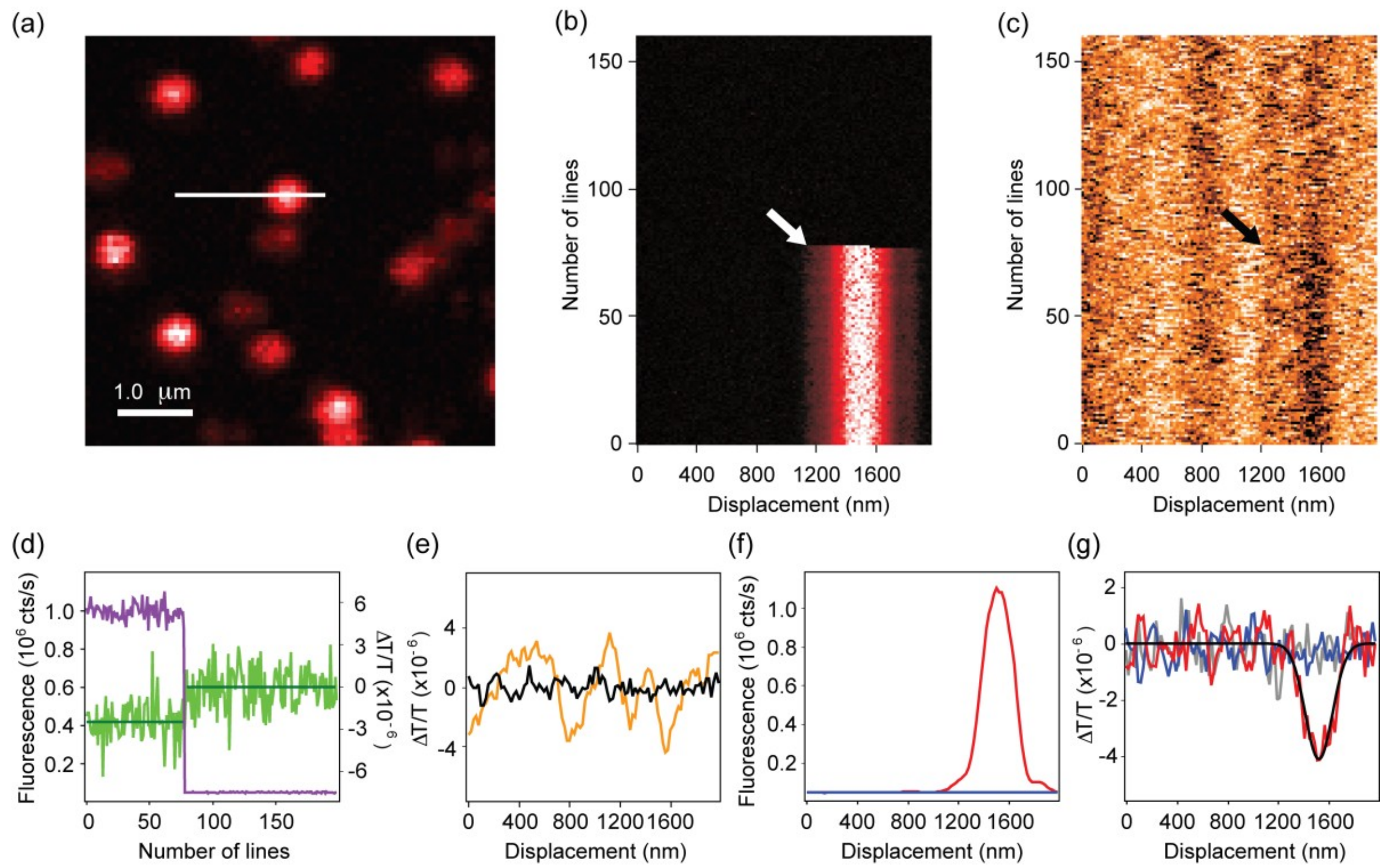

**Figure 2** (a) Fluorescence raster scan of TDI molecules in PMMA. (b) Fluorescence image constructed from repeated lateral scans of the molecule marked in (a). (c) Corresponding differential transmission image. Acquisition parameters are: 100 µW incident power, 1 ms pixel dwell time, 20 nm/pixel step size. (d) Fluorescence (violet) and absorption (green) cross sections from traces in (b) and (c) after averaging over 200 nm about the center of the molecule. (e) Lateral differential transmission scans averaged over 80 consecutive lines (orange) and subtraction of two such consecutive averages (black) in the absence of any molecule. (f) Averaged fluorescence line scans for the molecule in (b) before (red) and after (blue) photobleaching. (g) Corresponding differential transmission scans. Subtraction of the averaged scan in the off state from the averaged scan in the on state reveals a clear transmission dip (red). When both averages are chosen from the off (blue) or on (grey) states, no dip is visible.

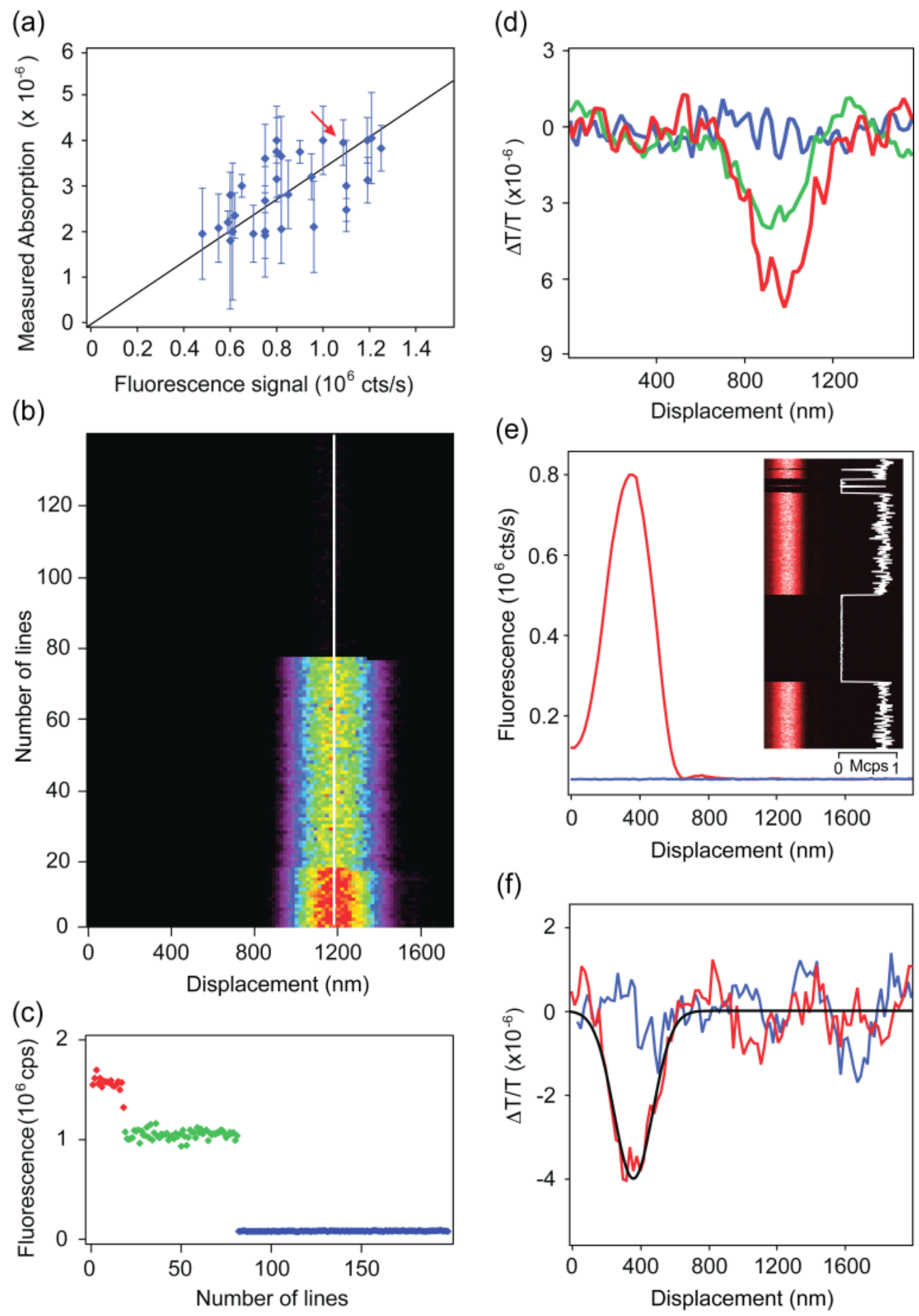

**Figure 3** (a) Correlation between absorption and fluorescence signals for 30 TDI molecules. Error bars reflect the shot noise and possible drift noise achieved for each single measurement. (b) Fluorescence line scan image for two close-lying molecules. (c) Averaged fluorescence intensity for the molecules in (b). (d) Corresponding differential transmission scans for two (green), one (red) and no (blue) emitting molecule. (e) Fluorescence line scan image for a molecule exhibiting long term fluorescence intermittency. The fluorescence intensity relative to a line cut is overlayed in white. (f) Corresponding differential transmission scans for the molecule in the on (red) and off (blue) states.

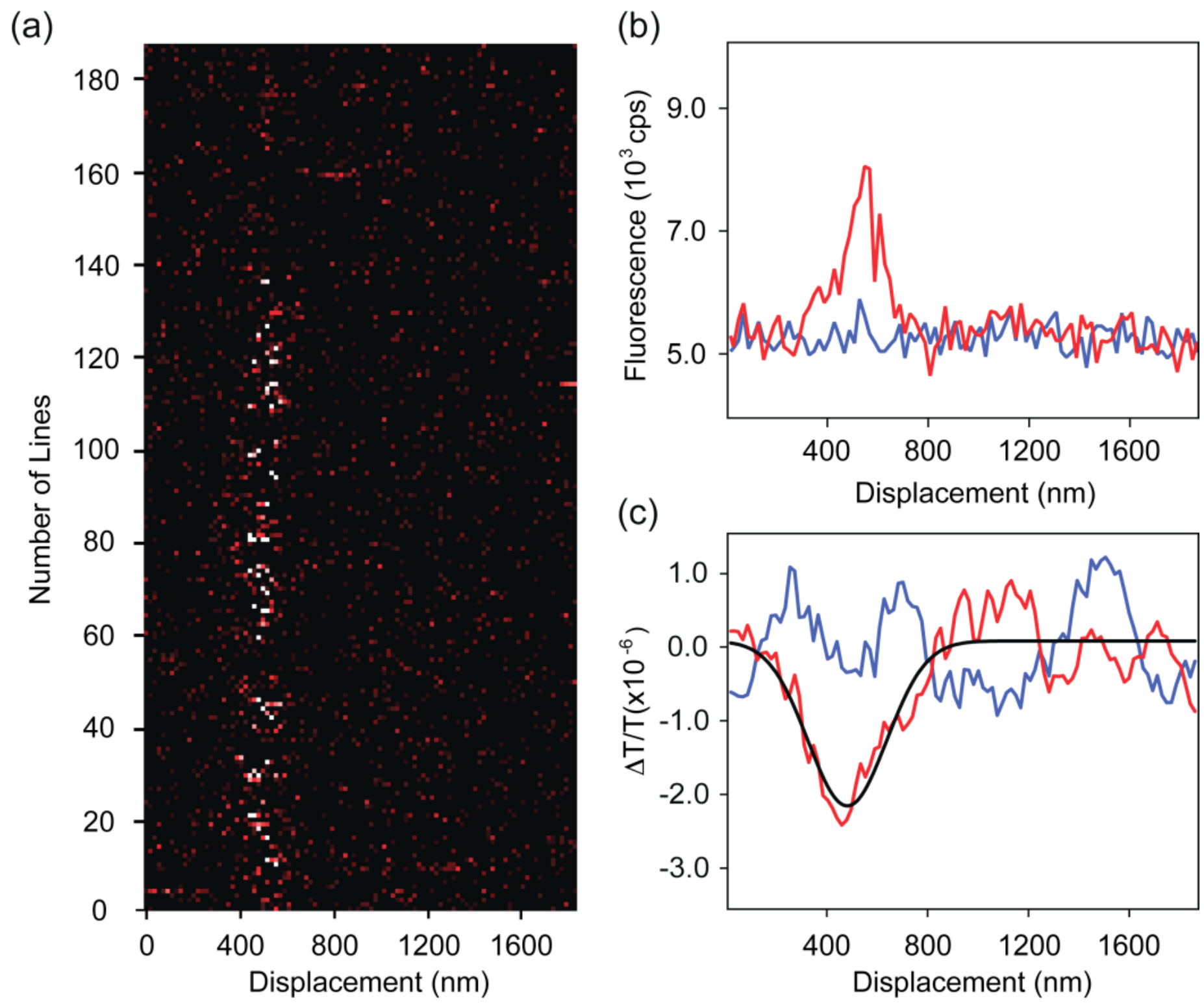

**Figure 4** (a) Fluorescence line scan image for a single Atto647N molecule covered by PVA. The incident power has been reduced to 10 μW to prevent rapid photobleaching. (b) Averaged fluorescence intensities before (red) and after (blue) photobleaching. (c) Corresponding differential transmission scans.